# Cross-polarized stimulated Brillouin scattering empowered photonics

Mingming Nie[1, *] Jonathan Musgrave[1] and Shu-Wei Huang[1, *]

[1]Department of Electrical, Computer and Energy Engineering, University of Colorado Boulder, Boulder, Colorado 80309, USA

[*]Corresponding author: mingming.nie@colorado.edu, shuwei.huang@colorado.edu

**Abstract:** This paper explores the integration of cross-polarized stimulated Brillouin scattering (XP-SBS) with Kerr and quadratic nonlinearities in lithium niobate (LN) to enhance photonic device performance. Three novel applications are demonstrated: (i) a reconfigurable stimulated Brillouin laser (SBL) with 0.7-Hz narrow linewidth and 40-nm tunability, enabled by XP-SBS's thermo-optic phase matching; (ii) an efficient coherent mode converter achieving 55% conversion efficiency via intracavity Brillouin-enhanced four-wave mixing; (iii) a Brillouin-quadratic laser and frequency comb operational in near-infrared and visible bands, benefiting from the interaction between XP-SBS and quadratic nonlinearity. These advancements promise significant improvements in photonic technologies, including narrow-linewidth laser, microcomb generation, and optical signal processing, paving the way for more robust and versatile applications.

**Introduction**

Stimulated Brillouin scattering (SBS) is a nonlinear coherent photon-phonon interaction that has been a subject of extensive study since its discovery over a century ago [1–3]. This process is characterized by high gain and narrow linewidth, making SBS pivotal in enhancing our understanding of light-matter interactions and has paved the way for innovative technologies including optical atomic clock [4], microwave photonics [5,6], high-resolution spectroscopy [7,8], optical gyroscope [9,10], optical signal processing [11–13], SBS microscopy [14–16] and quantum sensing [17,18].

Recent years see growing interest to study its fundamental physics and enhance SBS-based device performance by adding more degrees of freedom and combining other nonlinear optical phenomena. For example, it is demonstrated that optical angular momentum modes in ring-core optical fiber enable new control mechanism over SBS interactions [19] and coherent wavefront shaping in multimode fiber can be utilized to increase the SBS threshold for power scaling [20]. More importantly, SBS is introduced to solve a long-standing thermal instability challenge in the generation of dissipative Kerr soliton (DKS) microcomb, a ground-breaking microcomb technology with a remarkable breadth of demonstrated applications [21,22]. The interplay between the SBS and Kerr nonlinearity renders the Brillouin-DKS thermally self-stable, making the technology user-friendly and field-deployable [23–28].

Polarization is another degree of freedom that can be added to expand the SBS realm for more exciting applications. Despite the weak birefringence in optical fiber, cross-polarized SBS (XP-SBS) has been observed and utilized in Brillouin-DKS generation to fundamentally narrow the comb linewidth, lower the soliton timing jitter, and demonstrate the coveted turnkey microcomb [23]. In principle, use of anisotropic material such as lithium niobate (LN) can significantly increase the XP-SBS effect assuming the complex orientation-dependent optical, acoustic, and photoelastic properties are properly considered [29–32]. In practice, only until very recently has SBS in LN been studied theoretically and measured experimentally [33–36]. While XP-SBS in LN has also been observed spectroscopically [32,35], it remains an elusive question whether XP-SBS in LN can be made strong enough to be non-perturbative in interacting with other nonlinear phenomena and enhancing SBS-based device performances.

Besides the anisotropic property relevant to XP-SBS, LN is also non-centrosymmetric, meaning that it exhibits the second-order (quadratic) nonlinearity which is orders-of-magnitude stronger than the third-order (Kerr) nonlinearity. Thus, microcomb generation utilizing quadratic nonlinearity (quadratic microcomb), tends to have a lower pump threshold and higher pump-to-comb conversion efficiency compared with DKS microcomb [37] More importantly, quadratic nonlinearity provides a feasible solution to extend the microcomb wavelength to difficult-to-access ultraviolet and mid-infrared spectral ranges and enables fascinating applications including random number generation [38] and optical Ising machine [39]. Beyond these demonstrations, an intriguing question is whether XP-SBS can also be combined with quadratic nonlinearity to explore new opportunities and expand the already rich scope of quadratic microcomb [40].

This paper is the first attempt to address the two above-mentioned questions. We demonstrate three distinct photonic applications with unique features enabled by the strong XP-SBS and its interaction with Kerr and/or quadratic nonlinearities in LN (Fig. 1). (i) We devise the first reconfigurable stimulated Brillouin laser (SBL) with 0.7-Hz narrow fundamental linewidth over 40-nm broadband tunability and arbitrary SBL transverse mode selection. The reconfigurability and mode selection are made possible by the high thermo-optic phase matching tunability of XP-SBS in LN. (ii) We achieve the first efficient coherent mode convertor (polarization, transverse, and longitudinal modes) that utilizes intracavity Brillouin-enhanced four-wave mixing (BE-FWM) [41,42]. Mode conversion efficiency as high as 55% can be obtained thanks to the interplay between XP-SBS and co-polarized SBS (CP-SBS) in the quadruply resonant LN cavity. (iii) We realize the first Brillouin-quadratic laser and frequency comb at both the near-infrared and visible bands by combining XP-SBS and quadratic nonlinearity in periodically poled LN (PPLN). Strong and efficient interaction between the XP-SBS and quadratic nonlinearity is attainable due to the prohibition of the competing cascaded SBS process. Finally, switching between Brillouin-quadratic laser and frequency comb is achieved by changing the pump polarization.

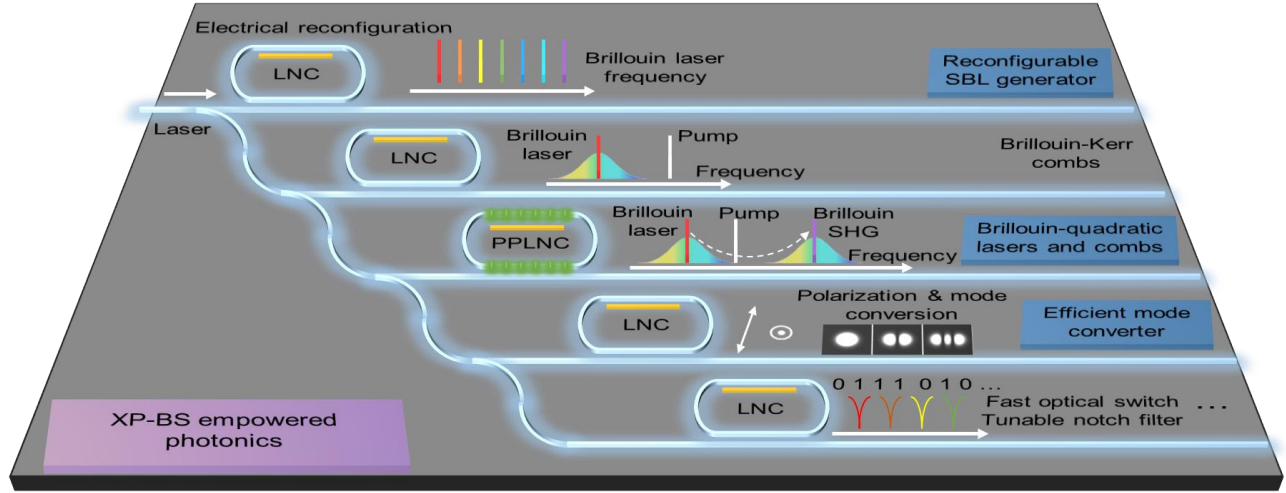

**Fig. 1. Schematic showing potential XP-SBS applications**. We experimentally demonstrate three distinct photonic applications here including reconfigurable SBL generator, efficient mode converter and Brillouin-quadratic lasers and combs.

## Results

**Reconfigurable SBL generator.** Achieving efficient excitation of SBL in a cavity requires a significant overlap between the Brillouin mode resonance and the Brillouin gain spectrum. The conventional approach to achieve this phase matching involves precisely controlling the cavity length within only a few micrometers [43], which is challenging especially for the monolithic cavity fabrication. Worse still, once the fabrication is finished, the SBL is not reconfigurable for broadband tunable operation due to the fixed cavity length and the resulted fixed phase matching condition, which is not preferred for field-deployable applications.

Our reconfigurable SBL generator consists of a birefringent free-space ring cavity within a z-cut-x-propagate LN crystal (see Methods and Supplementary Information for details). The Q factors are measured to be $2.136\times10^8$ and $2.147\times10^8$ for p and s polarizations at fundamental modes, respectively. The free spectral range (FSR) for the two polarizations are ~397 MHz and their FSR difference is measured to be 1.15 MHz.

Combining the XP-SBS gain, the SBL reconfigurability is fulfilled by adjusting the offset frequency between the orthogonal polarized mode families (Fig. 2a), which determines the SBS phase matching condition. Thanks to the FSR and thermo-optic coefficient difference, the offset frequency can be tuned via the LN crystal temperature with a coefficient of 188.8 MHz/°C (Fig. 2b), which means perfect phase matching condition can always be achieved within 2.1°C. Figure 2c depicts the dynamics of intracavity power of pump and SBL with slowly scanned crystal temperature with p-polarized pump and s-polarized SBL. In a critical range of LN crystal temperature, the intracavity pump energy can efficiently transfer to the SBL. Of note, reversing the polarizations for pump and SBL also works.

It is so convenient and efficient to tune the LN crystal temperature that perfect XP-SBS phase matching can always be achieved in our setup from 1540 nm to 1580 nm (Fig. 2d), which is limited by the amplifier operating wavelength and mirror coatings. In principle, the proposed birefringence-based tuning approach is universal for reconfigurable SBL at any wavelength. Besides, the suppression ratio (SR) between the SBL and the pump is as high as 48 dB (Fig. 2e) due to the counter-propagation and additional polarization-dependent suppression, which can benefit non-reciprocal devices based on SBS such as optical isolator [44]. In addition, by changing the crystal temperature, we are able to select the spatial mode of the XP-SBL participating in the photon-phonon interaction process (Fig. 2f), adding another degree of freedom for more exciting applications such as mode multiplexing.

We derive the Brillouin gain properties of bulk LN crystal for both the previously discussed XP-SBS case (p-s or s-p) and CP-SBS case (p-p or s-s) based on conventional cavity length tuning by studying the intracavity power dynamics (see Supplementary Information for details). As summarized in Table 1, XP-SBS process based on birefringence tuning has three overwhelming advantages: (i) flexible tunability for broadband SBL generation; (ii) SBL cascading inhibition and SBL power scaling; (iii) large SR between the SBL and the pump. Besides, the XP-SBS gain coefficient is more than three times that of the single-mode fiber (SMF). To the best of our knowledge, this is the first comprehensive study of SBS gain for bulk LN crystals.

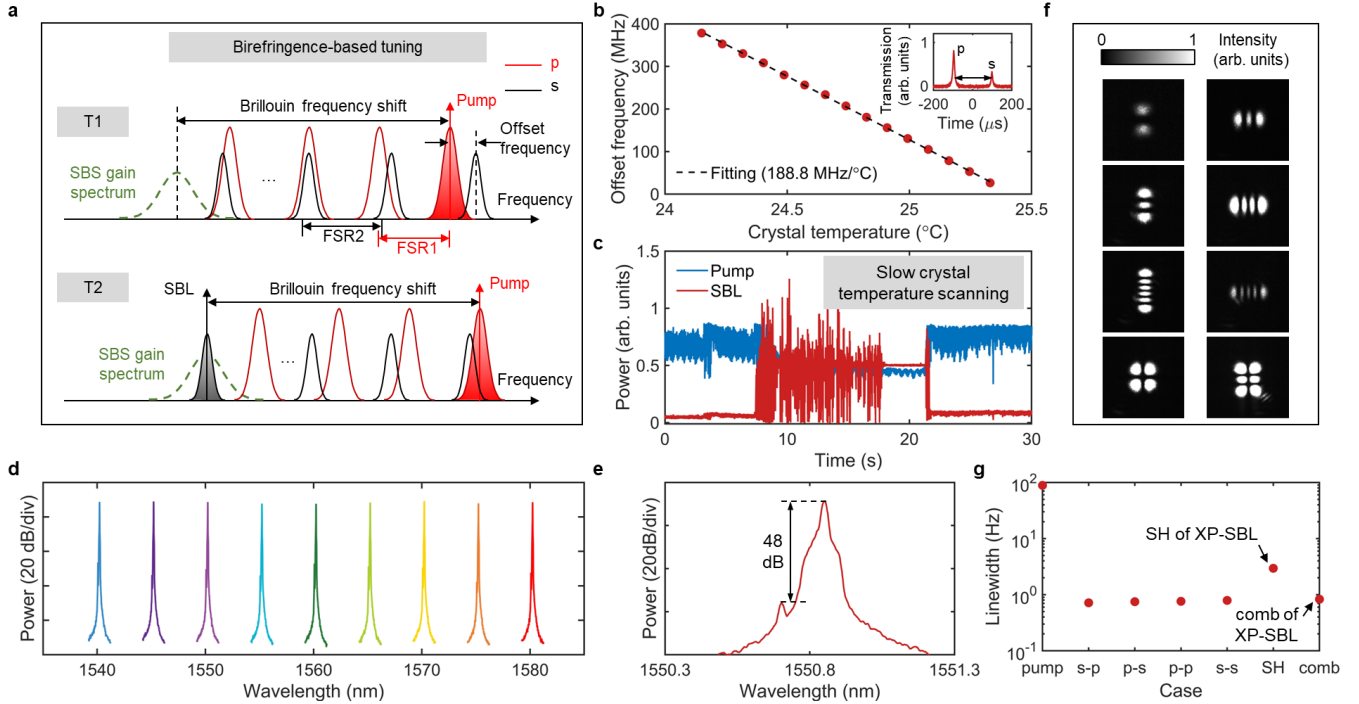

**Fig. 2. Reconfigurable SBL generator**. (a) Working principle. By tuning the crystal temperature from T1 to T2, correct offset frequency for perfect phase matching can be achieved. (b) Temperature-dependent offset frequency between two adjacent orthogonal polarized modes. (c) Intracavity power dynamics of pump and SBL with slowly scanned crystal temperature when the pump frequency is PDH locked to the cavity. (d) Measured spectra of the reconfigurable SBLs. (e) SBL spectrum with 48-dB SR. (f) High-order spatial mode profiles of the generated XP-SBL. (g) Measured fundamental linewidth.

Since the measured Brillouin gain bandwidth of ~10 MHz is ~10 times larger than the Brillouin mode cavity linewidth of 0.9 MHz, the phonons can quickly extract noise from the pump photons via the SBS process, resulting in low-noise SBL. The noise suppression factor is determined by $(1 + \Gamma_B/\gamma)^2$ [25], where $\Gamma_B$ is the Brillouin gain bandwidth and $\gamma$ is the Brillouin mode cavity linewidth. In Fig. 2g, we measure the fundamental linewidth for the pump and SBLs (see Methods and Supplementary Information for details). The noise suppressions from the pump are ~20 dB for all cases and lead to SBL fundamental linewidths of ~0.7 Hz, which agree well with the calculated results [45].

Table 1. Comparison of measured SBS gain properties and SBL generation at ~1550 nm

| Material & SBS type | Photoelastic coefficient | Theo. BFS[1] (GHz) | Exp. BFS (GHz) | Exp. BGC (cm/GW) | Exp. BGB (MHz) | Tunability (nm) | SR (dB) | CSBL generation |
|---|---|---|---|---|---|---|---|---|
| Bulk LN, XP-SBS (p-s[2], s-p) | $P_{41}$=0.151 | 18.435 | 18.441 | 15 | 9.49 | 40 | 48 | No |
| Bulk LN, CP-SBS (p-p) | $P_{31}$=0.09 | 18.755 | 18.759 | 6.8 | 8.37 | 0.289 | 27 | No[3] |
| Bulk LN, CP-SBS (s-s) | $p_{12}$=0.179 | 18.126 | 18.123 | 21 | 8.45 | 0.288 | 27 | Yes |
| SMF [46] | $p_{12}$=0.286 | ~11.1 | ~11.1 | 4.52 | 16 | - | - | - |

[1]Theo.: Theoretical, Exp.: Experimental, BFS: Brillouin frequency shift, BGC: Brillouin gain coefficient, BGB: Brillouin gain bandwidth, CSBL: cascaded SBL.
[2]p-s: p-polarized pump transfers energy to s-polarized SBL, s-p: s-polarized pump transfers energy to p-polarized SBL, p-p: p-polarized pump transfers energy to p-polarized SBL, s-s: s-polarized pump transfers energy to s-polarized SBL.
[3]CSBL is not found due to the insufficient pump power and low Brillouin gain coefficient.

**Efficient mode convertor.** The large XP-SBS gain and birefringence-based tunability can be applied for efficient mode conversion utilizing BE-FWM [41,42] in the cavity. In principle of BE-FWM, four light waves interact with each other through the Brillouin dynamic grating (BDG), which includes the BDG writing process with two lasers and the BDG reading-out process with another two lasers [41,42]. First, we write the BDG in the LN crystal by loading a s-polarized writing laser into the cavity and exciting a s-polarized SBL (Fig. 3a, also see Supplementary Information for details). The cavity length is carefully adjusted for perfectly phase matched CP-SBS (s-s) process with Brillouin frequency shift $\Omega_B$. Second, we load another p-polarized reading laser into the cavity with a power below the threshold of XP-SBL generation. The BDG information can be read out with the generation of a reflected s-polarized signal due to the extracted XP-SBS gain (p-s) from the written BDG. Since the reading laser co-propagates with the writing laser, the reflected signal is the Stokes light of the reading laser with a frequency shift equal to the acoustic frequency $\Omega_B$ of the written BDG. Thanks to the birefringent cavity, by changing the crystal temperature, the resonant frequency of s polarization in the cavity can be adjusted to match the frequency of the reflected s-polarized signal (Fig. 2a), leading to the cavity enhancement, large cavity output power for the reflected signal thus efficient mode conversion from the reading laser.

To obtain efficient reflection from the written BDG, the four lasers should satisfy the phase matching condition (Fig. 3b) [41,42] to share the same BDG and the frequency separation $\Delta\Omega$ between the writing and reading laser is determined by the birefringent condition [41,42]:

$$\frac{\Delta\Omega}{\omega_{write}} = \frac{\Delta n}{n_{write}} \quad (1)$$

where $\omega_{write}$ is the optical frequency of the writing laser, $n_{write}$ is the refractive index for the writing polarization and $\Delta n$ is the refractive index difference at the writing laser frequency. In our case, $n_{write} = n_e$ in the CP-SBS process with s to s polarization conversion and $n_{read} = (n_o + n_e)/2$ in the XP-SBS process with p to s polarization conversion, where $n_e$ and $n_o$ are the refractive index of extraordinary and ordinary light in the LN crystal.

As shown in Fig. 3c, when BDG is not written with writing laser turned off, no reflected signal is observed. With written BDG, we observe the reflected Stokes light of the reading laser. In addition, when the LN crystal temperature is adjusted, the reflected signal can be resonant in the cavity with significant power enhancement. The frequency separation between the writing and reading lasers is measured to be as large as 27.7 nm, which agrees well with the calculated one of 27 nm. Figures 3d and 3e show the output power of the resonant reflected signal against different reading and writing laser powers, respectively. Maximum output power of 120 mW and conversion efficiency of 55% from reading laser to the reflected signal are achieved due to the large XP-SBS gain and birefringence-tuned cavity enhancement, which agree very well with the simulated results (dashed lines in Figs. 3d and 3e, also see Supplementary Information for details). Similarly, we can also select the spatial mode profile of the resonant reflected signal (Fig. 3f) thanks to the flexible birefringence tuning.

To the best of our knowledge, this is the first demonstration of efficient mode conversion utilizing intracavity BE-FWM. With only two pump lasers, another two lasers are generated, and all four lasers are resonant in the cavity thanks to the birefringence tuning in the anisotropic cavity. Mode properties such as polarization, spatial mode profile and optical frequency are all converted to other ones with high efficiency. Of note, same polarization conversion as well as frequency up-conversion with anti-Stokes generation of the reading laser [47] are possible by changing the BE-FWM configuration [48]. Apart from the efficient mode conversion, the demonstrated intracavity BE-FWM are believed to benefit applications such as optical sensing [49], all-optical signal processing [11], all-optical delay [50], microwave photonic filter [51], ultrahigh resolution optical spectrometry [52] and so on.

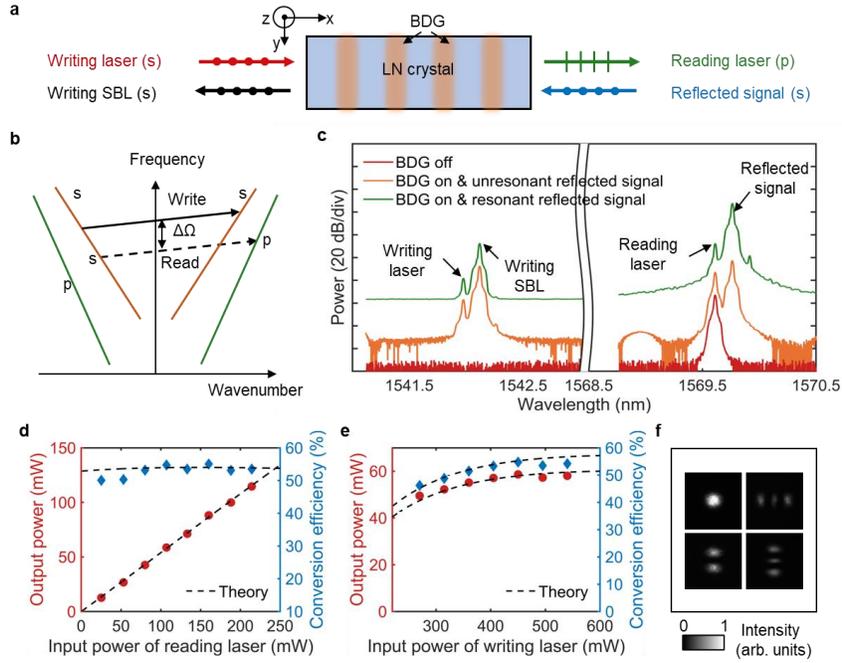

**Fig. 3. Efficient mode conversion with intracavity BE-FWM.** (a) Schematic of BE-FWM process. (b) Phase matching diagram for BE-FWM process. (c) Optical spectra showing the BE-FWM process. (d) Output power of the resonant reflected signal by changing the input power of reading laser. The input power of the writing laser is 520 mW. (e) Output power of the resonant reflected signal by changing the input power of writing laser. The input power of the reading laser is 120 mW. (f) Spatial mode profiles of the resonant reflected signal.

**Efficient visible laser and comb generator.** Figure 4 shows efficient visible laser and comb generation by combining the XP-SBS effect and polarization-dependent quadratic effect in a PPLN crystal. In Fig. 4a, the p-polarized SBL from s-polarized pump can efficiently generate second harmonic (SH) signal at 775 nm through type-I phase matching condition ($o+o\rightarrow e$) with a nonlinear coefficient $d_{eff}$=2.8 pm/V. The output power of the single-frequency 775 nm is as high as 217 mW at a pump power of 1800 mW (Fig. 4b), corresponding to a pump conversion efficiency of 12%. Higher output power is expected by optimizing the cavity parameters such as the output coupling [53]. The fundamental linewidth of the generated visible laser (Fig. 4c) is measured to be 2.8 Hz (Fig. 2g), which is 6 dB larger than the p-polarized SBL as expected from theory.

In Fig. 4d, we first generate s-polarized SBL from p-polarized pump and then utilize the type-0 phase matching condition ($e+e\rightarrow e$) with a larger $d_{eff}$=15.2 pm/V for quadratic comb generation due to the cascading between the four quadratic processes, such as SH generation (SHG), optical parametric generation (OPG), sum frequency generation (SFG) and difference frequency generation (DFG) [54]. In Figs. 4e and 4f, we plot the Turing pattern generation for both bands. We also observe the chaotic comb generation in the Supplementary Information. To the best of our knowledge, this is the first demonstration of visible Brillouin-quadratic laser and comb generation.

Of note, the pump experiences no $\chi^{(2)}$ nonlinearity but only efficiently transfers the energy to the XP-SBL. In addition, compared to the visible SBL generation in microresonators whose noise suppression is limited by the large cavity loss [55], the Brillouin-quadratic process can take full advantage of the narrow linewidth at near-infrared band. Therefore, the Brillouin-quadratic process provides a feasible and efficient route toward visible laser and comb generation with ultralow noise (Fig. 2g). Similar to our previously reported Brillouin-DKS comb generation [24,25,56], we anticipate feasible Brillouin-quadratic soliton comb generation [57] with ultralow fundamental linewidth as well as ultralow timing jitter through dispersion engineering, for example in Lithium niobate on insulator (LNOI) microresonators, fulfilling the long-sought-after high-coherence laser sources across a range of applications.

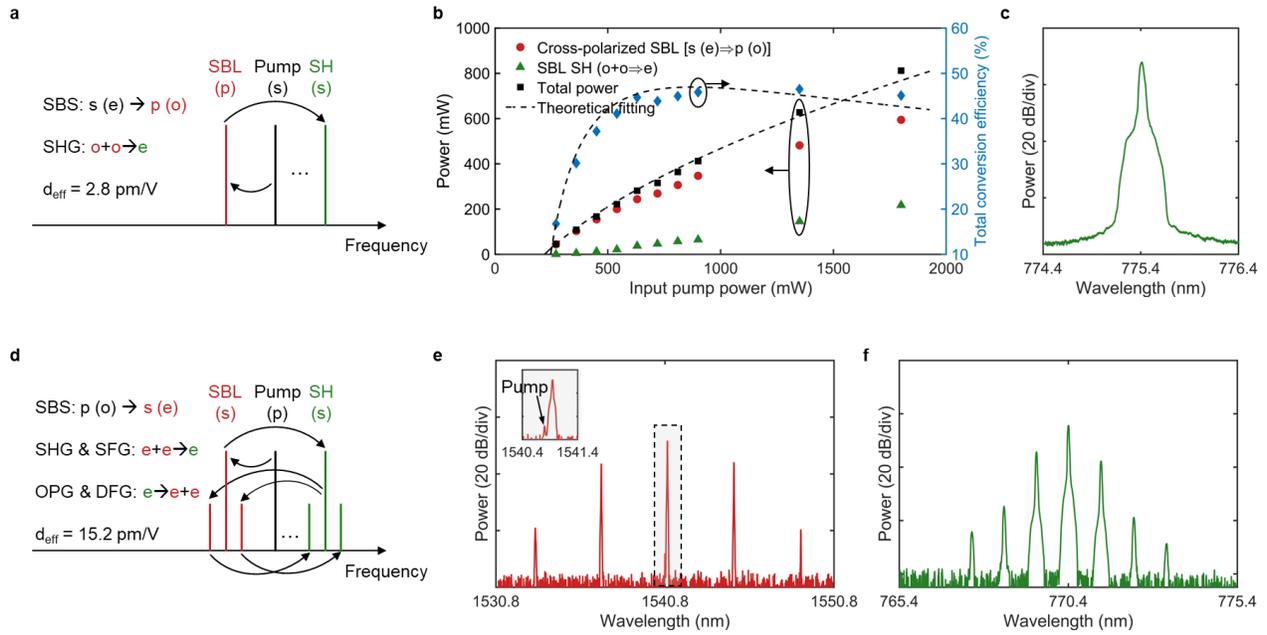

**Fig. 4. Efficient visible laser and comb generation.** (a)(b)(c) SBS process: s-p; SHG process: type-I phase matching, $o+o\rightarrow e$. (a) Schematic of the whole processes. (b) Output power of 1st order SBL and its SH signal against input pump power. Total power and corresponding conversion efficiency are also plotted. (c) Spectrum of SBL SH signal. (d)(e)(f) SBS process: p-s; quadratic processes: type-0 phase matching, $e+e\rightarrow e$ or $e\rightarrow e+e$. (d) Schematic of the whole processes. (e) Spectra of Turing pattern state centered at 1540.8 nm. Inset is zoomed in from the dashed box showing large SR between pump and XP-SBL. (f) Spectra of Turing pattern state centered at 770.4 nm.

## Discussion

Besides the thermal-optic effect, other various effects in LN crystals such as electro-optic (EO) effect, acousto-optic (AO) effect, photorefractive effect [31] and so on can be used for faster birefringence tuning. In addition, there are other birefringent crystals possessing non-diagonal photoelastic coefficients and XP-SBS gain, such as $BiFeO_3$ crystal [32], $LiTaO_3$ crystal [58]. Moreover, XP-SBS can be extended to on-chip platforms for further performance improvement. For example, by introducing surface acoustic wave in a nano-dimensional waveguide with well-confined optical and acoustic modes, the emergent thin-film LNOI shows strong XP-SBS gain [33–35] and the potential of a very promising platform for integrated Brillouin photonic devices and photonic circuits for optical information processing engine with low power consumption. The utilization of on-chip XP-SBS gain, including the optically stimulated Brillouin gain demonstrated here and the electrically driven Brillouin gain, can contribute to more powerful devices, for example on-chip isolators [44], fast optical switch [59], tunable notch filter [12,13] and so on.

## Methods

**Experimental details.** The free-space cavity is made from four-mirror bow-tie cavity with a FSR of 397 MHz. Three reflective mirrors (M1-M3) are coated with high reflectivity >99.95% from 1530 nm to 1580 nm. The output coupler is coated with high reflectivity of 1.03% from 1530 nm to 1580 nm. The four mirrors are coated with high transmission >95% from 760 nm to 790 nm. Both end facets of the 25-mm long 5% MgO-doped PPLN crystal are coated with high transmission at 1550 nm and 775 nm (reflectivity of 0.13% at 1550 nm and 0.09% at 780 nm). The cavity linewidth and Q factor are measured through a frequency-calibrated MZI with a FSR of 0.998 MHz. The offset frequency between the two adjacent polarized modes (within one cavity FSR) in Fig. 2b is measured via the same MZI with scanned pump frequency, and the pump is linearly polarized at 30° with respect to the p polarization. The input power and output power are all measured from the output coupler.

The PPLN crystal has two poling periods, 20.6 μm for type-I phase matching at ~1550 nm and 19.1 μm for type-0 phase matching at ~1541 nm. The PPLN crystal is temperature controlled with a resolution of 10 mK and the whole cavity is enclosed in a plastic box to prevent heat exchange with the outside air. When the temperature of PPLN deviates from the phase matching condition for cascaded quadratic processes, efficient XP-SBS process is still possible since there are multiple temperature points for perfect SBS phase matching. And XP-SBS phase matching is more sensitive to the temperature change compared with the phase matching for quadratic processes.

The pump is locked to the free-space cavity for efficient SBL generation via the powerful PDH technique. The phase modulation frequency is 1 MHz and the PDH signal is demodulated by the pump output from M2. The low-pass filter used in the PDH signal demodulation is 80 kHz. The modulation voltage of the phase modulator is chosen to be low without perturbing the SBL generation.

**LN material properties.** The photoelastic tensor of LN is given by the 6×6 matrix [31]

$$p_{ijkl} = \begin{bmatrix} p_{11} & p_{12} & p_{13} & p_{14} & 0 & 0 \\ p_{12} & p_{11} & p_{13} & -p_{14} & 0 & 0 \\ p_{31} & p_{31} & p_{33} & 0 & 0 & 0 \\ p_{41} & -p_{41} & 0 & p_{44} & 0 & 0 \\ 0 & 0 & 0 & 0 & p_{44} & p_{41} \\ 0 & 0 & 0 & 0 & p_{14} & \frac{p_{11}-p_{12}}{2} \end{bmatrix}, \quad (2)$$

where $p_{12}$=0.09, $p_{31}$=0.179 and $p_{41}$=0.151 [58] determine the SBS gain coefficients for p to p polarization conversion, s to s polarization conversion and cross-polarized conversion (p to s or s to p polarization conversion) [32], respectively.

**SBS gain properties calculation.** Assuming both pump and SBL mode are with fundamental mode (HG$_{00}$ mode), the SBL output $P_B$ in a free-space cavity is given by [53]

$$P_B = \sigma_S \left( \sqrt{\frac{P_{in}}{P_{th}}} - 1 \right), \quad (3)$$

where

$$P_{th} = \frac{\pi w_0^2 \left[1 - R(1-L)\right]^3}{4 g_B l_{eff} \eta_{oo} (1-R)}, \quad (4)$$

$$\sigma_S = \frac{\pi w_0^2 (1-R)\left[1 - R(1-L)\right]}{g_B l_{eff}}, \quad (5)$$

$P_{in}$ is the input pump power, $w_0$ =50 μm is the beam radius inside the crystal, $R$ =0.9897 is the reflectivity of the OC mirror, $L$ =0.004 is the other total passive intracavity loss, $\eta_{oo}$ is the optical mode overlap between the pump and SBL mode (we use 0.9 for XP case due to the astigmatism and use 0.95 for CP cases), $g_B$ is the Brillouin gain coefficient and $l_{eff}$ is the effective SBS interaction length which is determined by

$$l_{eff} = b \arctan\left(\frac{l}{b}\right), \quad (6)$$

where $l$ =25 mm is the LN crystal length and $b$ is the confocal parameter given by

$$b = \frac{2\pi n w_0^2}{\lambda_p}, \quad (7)$$

where $n$ is refractive index and $\lambda_p$ =1.55 μm is the pump wavelength. By substituting the experimental input pump and output SBL power into Eq. (3) and conduct a fitting, we can obtain the Brillouin gain coefficient.

With the Brillouin gain coefficient, we can calculate the spontaneous Brillouin gain bandwidth through [60]

$$g_B = \eta_{oa} \frac{4\pi n^8 p^2}{c \rho v_B \Delta v_B \lambda_p^3}, \quad (8)$$

where $\eta_{oa}$ =0.95 is the overlap integral between the optical mode and acoustic mode induced density change, $p$ is the photoelastic coefficient characterizing the SBS process, $c$ is the speed of light, $\rho$ =4647 kg/m$^3$ is the density, $v_B$ is Brillouin frequency shift and $\Delta v_B$ is Brillouin gain bandwidth.

The Brillouin frequency shift can be calculated through

$$\Omega_B = \frac{2n\upsilon_s}{\lambda_p}, \quad (9)$$

where $\upsilon_s$ =6572 m/s is the speed of longitudinal acoustic wave in LN crystals [33,61]. In the case of co-polarized case with p to p polarization conversion, $n = n_o = 2.2118$ ($n_o$ is the refractive index of ordinary light in LN crystals) and $\Omega_B$ is calculated to be 18.755 GHz. In the case of co-polarized case with s to s polarization conversion, $n = n_e = 2.1376$ ($n_e$ is the refractive index of extraordinary light in LN crystals) and $\Omega_B$ is calculated to be 18.126 GHz. In the case of cross-polarized case, $n = (n_o + n_e)/2 = 2.1747$ and $\Omega_B$ is calculated to be 18.435 GHz.

**Measurement of fundamental linewidth.** Two self-heterodyne frequency discriminators using a fiber based unbalanced MZI and a balanced photodetector (BPD) is employed to measure the laser phase noise and fundamental linewidth for SBL and its SH signal, respectively. For both discriminators, one arm of the unbalanced MZI is made of 250-m-long single mode fiber, while the other arm consists of an acousto-optic frequency shifter with frequency shift of 200 MHz and a polarization controller for high-voltage output. The FSRs of the unbalanced MZI are 0.85 MHz and MHz for 1550 nm and 775 nm, respectively. The two 50:50 outputs of the unbalanced MZI are connected to a BPD with a bandwidth of 400 MHz to reduce the impact of detector intensity fluctuations. The balanced output is then analyzed by a phase noise analyzer (NTS-1000A, RDL). We do not measure the visible comb linewidth of the generated Turing pattern due to the lack of an optical filter. According to the measured results at ~1545 nm, the visible comb linewidth should have the same fundamental linewidth with its pump (SH of the generated SBL), which is 6 dB larger than the generated SBL.

**Data availability**
All data generated or analyzed during this study are available within the paper and its Supplementary Information. Further source data will be made available on request.

**Code availability**
The analysis codes will be available on request.

**Acknowledgments**
M.N. J.M. and S.W.H. acknowledge the support from the National Science Foundation (ECCS2048202), Office of Naval Research (N00014-22-1-2224), and National Institute of Biomedical Imaging and Bioengineering (REB029541A).

**Author contributions**
M.N. conceived the idea and designed the experiment. M.N. and J. M. performed the experiment and simulation. M.N. J. M. and S.W.H. conducted the data analysis and wrote the manuscript. S.W.H. led and supervised the whole project. All authors contributed to the discussion and revision of the manuscript.

**Competing interests**
M.N. and S.W.H. are the inventors of a provisional patent application, filed by the University of Colorado Boulder, about the reconfigurable Brillouin lasers and Brillouin-quadratic lasers and frequency combs approach.

## References

[1] M. Merklein, I.V. Kabakova, A. Zarifi, B.J. Eggleton, 100 years of Brillouin scattering: Historical and future perspectives, Appl. Phys. Rev. 9 (2022). https://doi.org/10.1063/5.0095488.

[2] B.J. Eggleton, C.G. Poulton, P.T. Rakich, M. Steel, G. Bahl, others, Brillouin integrated photonics, Nat. Photonics 13 (2019) 664–677. https://doi.org/10.1038/s41566-019-0498-z.

[3] C. Wolff, M. Smith, B. Stiller, C. Poulton, Brillouin scattering—theory and experiment: tutorial, JOSA B 38 (2021) 1243–1269.

[4] W. Loh, J. Stuart, D. Reens, C.D. Bruzewicz, D. Braje, J. Chiaverini, P.W. Juodawlkis, J.M. Sage, R. McConnell, Operation of an optical atomic clock with a Brillouin laser subsystem, Nature 588 (2020) 244–249. https://doi.org/10.1038/s41586-020-2981-6.

[5] S. Gundavarapu, G.M. Brodnik, M. Puckett, T. Huffman, D. Bose, R. Behunin, J. Wu, T. Qiu, C. Pinho, N. Chauhan, others, Sub-hertz fundamental linewidth photonic integrated Brillouin laser, Nat. Photonics 13 (2019) 60–67. https://doi.org/10.1038/s41566-018-0313-2.

[6] J. Li, H. Lee, K.J. Vahala, Microwave synthesizer using an on-chip Brillouin oscillator, Nat. Commun. 4 (2013) 1–7. https://doi.org/10.1038/ncomms3097.

[7] A. Cygan, D. Lisak, P. Morzyński, M. Bober, M. Zawada, E. Pazderski, R. Ciury\lo, Cavity mode-width spectroscopy with widely tunable ultra narrow laser, Opt. Express 21 (2013) 29744–29754. https://doi.org/10.1364/OE.21.029744.

[8] A. Jadbabaie, N.H. Pilgram, J. K\los, S. Kotochigova, N.R. Hutzler, Enhanced molecular yield from a cryogenic buffer gas beam source via excited state chemistry, New J. Phys. 22 (2020) 022002. https://doi.org/10.1088/1367-2630/ab6eae.

[9] Y.-H. Lai, M.-G. Suh, Y.-K. Lu, B. Shen, Q.-F. Yang, H. Wang, J. Li, S.H. Lee, K.Y. Yang, K. Vahala, Earth rotation measured by a chip-scale ring laser gyroscope, Nat. Photonics (2020) 1–5. https://doi.org/10.1038/s41566-020-0588-y.

[10] Y.-H. Lai, Y.-K. Lu, M.-G. Suh, Z. Yuan, K. Vahala, Observation of the exceptional-point-enhanced Sagnac effect, Nature 576 (2019) 65–69. https://doi.org/10.1038/s41586-019-1777-z.

[11] M. Santagiustina, S. Chin, N. Primerov, L. Ursini, L. Thévenaz, All-optical signal processing using dynamic Brillouin gratings, Sci. Rep. 3 (2013) 1594. https://doi.org/10.1038/srep01594.

[12] S. Gertler, N.T. Otterstrom, M. Gehl, A.L. Starbuck, C.M. Dallo, A.T. Pomerene, D.C. Trotter, A.L. Lentine, P.T. Rakich, Narrowband microwave-photonic notch filters using Brillouin-based signal transduction in silicon, Nat. Commun. 13 (2022) 1947. https://doi.org/10.1038/s41467-022-29590-0.

[13] M. Garrett, Y. Liu, M. Merklein, C.T. Bui, C.K. Lai, D.-Y. Choi, S.J. Madden, A. Casas-Bedoya, B.J. Eggleton, Integrated microwave photonic notch filter using a heterogeneously integrated Brillouin and active-silicon photonic circuit, Nat. Commun. 14 (2023) 7544. https://doi.org/10.1038/s41467-023-43404-x.

[14] G. Scarcelli, S.H. Yun, Confocal Brillouin microscopy for three-dimensional mechanical imaging, Nat. Photonics 2 (2008) 39–43.

[15] I. Remer, R. Shaashoua, N. Shemesh, A. Ben-Zvi, A. Bilenca, High-sensitivity and high-specificity biomechanical imaging by stimulated Brillouin scattering microscopy, Nat. Methods 17 (2020) 913–916.

[16] F. Yang, C. Bevilacqua, S. Hambura, A. Neves, A. Gopalan, K. Watanabe, M. Govendir, M. Bernabeu, J. Ellenberg, A. Diz-Muñoz, others, Pulsed stimulated Brillouin microscopy enables high-sensitivity mechanical imaging of live and fragile biological specimens, Nat. Methods 20 (2023) 1971–1979.

[17] C.L. Degen, F. Reinhard, P. Cappellaro, Quantum sensing, Rev. Mod. Phys. 89 (2017) 035002. https://doi.org/10.1103/RevModPhys.89.035002.

[18] H. Fan, S. Kumar, J. Sedlacek, H. Kübler, S. Karimkashi, J.P. Shaffer, Atom based RF electric field sensing, J. Phys. B At. Mol. Opt. Phys. 48 (2015) 202001. https://doi.org/10.1088/0953-4075/48/20/202001.

[19] A.P. Greenberg, Z. Ma, S. Ramachandran, Angular momentum driven dynamics of stimulated Brillouin scattering in multimode fibers, Opt. Express 30 (2022) 29708–29721.

[20] C.-W. Chen, L.V. Nguyen, K. Wisal, S. Wei, S.C. Warren-Smith, O. Henderson-Sapir, E.P. Schartner, P. Ahmadi, H. Ebendorff-Heidepriem, A.D. Stone, others, Mitigating stimulated Brillouin scattering in multimode fibers with focused output via wavefront shaping, Nat. Commun. 14 (2023) 7343.

[21] T.J. Kippenberg, A.L. Gaeta, M. Lipson, M.L. Gorodetsky, Dissipative Kerr solitons in optical microresonators, Science 361 (2018) eaan8083. https://doi.org/10.1126/science.aan8083.

[22] A.L. Gaeta, M. Lipson, T.J. Kippenberg, Photonic-chip-based frequency combs, Nat. Photonics 13

[23] M. Nie, J. Musgrave, K. Jia, J. Bartos, S. Zhu, Z. Xie, S.-W. Huang, Turnkey photonic flywheel in a microresonator-filtered laser, Nat. Commun. 15 (2024) 55. https://doi.org/10.1038/s41467-023-44314-8.

[24] M. Nie, K. Jia, Y. Xie, S. Zhu, Z. Xie, S.-W. Huang, Synthesized spatiotemporal mode-locking and photonic flywheel in multimode mesoresonators, Nat. Commun. 13 (2022) 1–9. https://doi.org/10.1038/s41467-022-34103-0.

[25] K. Jia, X. Wang, D. Kwon, J. Wang, E. Tsao, H. Liu, X. Ni, J. Guo, M. Yang, X. Jiang, others, Photonic flywheel in a monolithic fiber resonator, Phys. Rev. Lett. 125 (2020) 143902.

[26] Y. Bai, M. Zhang, Q. Shi, S. Ding, Y. Qin, Z. Xie, X. Jiang, M. Xiao, Brillouin-Kerr soliton frequency combs in an optical microresonator, Phys. Rev. Lett. 126 (2021) 063901.

[27] I.H. Do, D. Kim, D. Jeong, D. Suk, D. Kwon, J. Kim, J.H. Lee, H. Lee, Self-stabilized soliton generation in a microresonator through mode-pulled Brillouin lasing, Opt. Lett. 46 (2021) 1772–1775.

[28] M. Zhang, S. Ding, X. Li, K. Pu, S. Lei, M. Xiao, X. Jiang, Strong interactions between solitons and background light in Brillouin-Kerr microcombs, Nat. Commun. 15 (2024) 1661.

[29] G. Lin, J. Tian, T. Sun, Q. Song, Y.K. Chembo, Hundredfold increase of stimulated Brillouin-scattering bandwidth in whispering-gallery mode resonators, Photonics Res. 11 (2023) 917–924. https://doi.org/10.1364/PRJ.484727.

[30] J. Tian, G. Lin, Theoretical analysis of the influence of crystalline orientation on Brillouin gain of whispering gallery mode resonators, JOSA B 41 (2024) 712–719. https://doi.org/10.1364/JOSAB.509176.

[31] R. Weis, T. Gaylord, Lithium niobate: Summary of physical properties and crystal structure, Appl. Phys. A 37 (1985) 191–203. https://doi.org/10.1007/BF00614817.

[32] M. Lejman, G. Vaudel, I.C. Infante, I. Chaban, T. Pezeril, M. Edely, G.F. Nataf, M. Guennou, J. Kreisel, V.E. Gusev, others, Ultrafast acousto-optic mode conversion in optically birefringent ferroelectrics, Nat. Commun. 7 (2016) 1–10. https://doi.org/10.1038/ncomms12345.

[33] C.C. Rodrigues, R.O. Zurita, T.P. Alegre, G.S. Wiederhecker, Stimulated Brillouin scattering by surface acoustic waves in lithium niobate waveguides, JOSA B 40 (2023) D56–D63. https://doi.org/10.1364/JOSAB.482656.

[34] K. Ye, H. Feng, Y. Klaver, A. Keloth, A. Mishra, C. Wang, D. Marpaung, Surface acoustic wave stimulated Brillouin scattering in thin-film lithium niobate waveguides, ArXiv Prepr. ArXiv231114697 (2023).

[35] C.C. Rodrigues, N.J. Schilder, R.O. Zurita, L.S. Magalhães, A. Shams-Ansari, T.P. Alegre, M. Lončar, G.S. Wiederhecker, On-Chip Backward Stimulated Brillouin Scattering in Lithium Niobate Waveguides, ArXiv Prepr. ArXiv231118135 (2023).

[36] Y.-H. Yang, J.-Q. Wang, Z.-X. Zhu, X.-B. Xu, Q. Zhang, J. Lu, Y. Zeng, C.-H. Dong, L. Sun, G.-C. Guo, others, Stimulated Brillouin interaction between guided phonons and photons in a lithium niobate waveguide, Sci. China Phys. Mech. Astron. 67 (2024) 214221.

[37] A.W. Bruch, X. Liu, Z. Gong, J.B. Surya, M. Li, C.-L. Zou, H.X. Tang, Pockels soliton microcomb, Nat. Photonics 15 (2021) 21–27.

[38] Y. Liu, Q. Zhao, M.-H. Li, J.-Y. Guan, Y. Zhang, B. Bai, W. Zhang, W.-Z. Liu, C. Wu, X. Yuan, others, Device-independent quantum random-number generation, Nature 562 (2018) 548–551.

[39] T. Inagaki, Y. Haribara, K. Igarashi, T. Sonobe, S. Tamate, T. Honjo, A. Marandi, P.L. McMahon, T. Umeki, K. Enbutsu, O. Tadanaga, H. Takenouchi, K. Aihara, K. Kawarabayashi, K. Inoue, S. Utsunomiya, H. Takesue, A coherent Ising machine for 2000-node optimization problems, Science 354 (2016) 603–606. https://doi.org/10.1126/science.aah4243.

[40] M. Nie, Y. Xie, B. Li, S.-W. Huang, Photonic frequency microcombs based on dissipative Kerr and quadratic cavity solitons, Prog. Quantum Electron. (2022) 100437. https://doi.org/10.1016/j.pquantelec.2022.100437.

[41] K.Y. Song, W. Zou, Z. He, K. Hotate, All-optical dynamic grating generation based on Brillouin scattering in polarization-maintaining fiber, Opt. Lett. 33 (2008) 926–928. https://doi.org/10.1364/OL.33.000926.

[42] R. Pant, E. Li, C.G. Poulton, D.-Y. Choi, S. Madden, B. Luther-Davies, B.J. Eggleton, Observation of Brillouin dynamic grating in a photonic chip, Opt. Lett. 38 (2013) 305–307. https://doi.org/10.1364/OL.38.000305.

[43] J. Li, H. Lee, T. Chen, K.J. Vahala, Characterization of a high coherence, Brillouin microcavity laser on silicon, Opt. Express 20 (2012) 20170–20180.

[44] C.G. Poulton, R. Pant, A. Byrnes, S. Fan, M. Steel, B.J. Eggleton, Design for broadband on-chip isolator using stimulated Brillouin scattering in dispersion-engineered chalcogenide waveguides, Opt. Express 20 (2012) 21235–21246. https://doi.org/10.1364/OE.20.021235.

[45] B. Li, Z. Yuan, W. Jin, L. Wu, J. Guo, Q.-X. Ji, A. Feshali, M. Paniccia, J.E. Bowers, K.J. Vahala, High-coherence hybrid-integrated 780 nm source by self-injection-locked second-harmonic generation in a

high-Q silicon-nitride resonator, Optica 10 (2023) 1241–1244. https://doi.org/10.1364/OPTICA.498391.

[46] K.S. Abedin, Observation of strong stimulated Brillouin scattering in single-mode As 2 Se 3 chalcogenide fiber, Opt. Express 13 (2005) 10266–10271. https://doi.org/10.1364/OPEX.13.010266.

[47] Y. Dong, L. Chen, X. Bao, Characterization of the Brillouin grating spectra in a polarization-maintaining fiber, Opt. Express 18 (2010) 18960–18967. https://doi.org/10.1364/OE.18.018960.

[48] H. Zhang, Y. Dong, Advances in Brillouin dynamic grating in optical fibers and its applications, Prog. Quantum Electron. 87 (2023) 100440. https://doi.org/10.1016/j.pquantelec.2022.100440.

[49] K.Y. Song, W. Zou, Z. He, K. Hotate, Optical time-domain measurement of Brillouin dynamic grating spectrum in a polarization-maintaining fiber, Opt. Lett. 34 (2009) 1381–1383. https://doi.org/10.1364/OL.34.001381.

[50] K.Y. Song, K. Lee, S.B. Lee, Tunable optical delays based on Brillouin dynamic grating in optical fibers, Opt. Express 17 (2009) 10344–10349. https://doi.org/10.1364/OE.17.010344.

[51] J. Sancho, N. Primerov, S. Chin, Y. Antman, A. Zadok, S. Sales, L. Thévenaz, Tunable and reconfigurable multi-tap microwave photonic filter based on dynamic Brillouin gratings in fibers, Opt. Express 20 (2012) 6157–6162. https://doi.org/10.1364/OE.20.006157.

[52] Y. Dong, T. Jiang, L. Teng, H. Zhang, L. Chen, X. Bao, Z. Lu, Sub-MHz ultrahigh-resolution optical spectrometry based on Brillouin dynamic gratings, Opt. Lett. 39 (2014) 2967–2970. https://doi.org/10.1364/OL.39.002967.

[53] D. Jin, Z. Bai, M. Li, X. Yang, Y. Wang, R.P. Mildren, Z. Lu, Modeling and characterization of high-power single frequency free-space Brillouin lasers, Opt. Express 31 (2023) 2942–2955. https://doi.org/10.1364/OE.476759.

[54] F. Leo, T. Hansson, I. Ricciardi, M. De Rosa, S. Coen, S. Wabnitz, M. Erkintalo, Walk-off-induced modulation instability, temporal pattern formation, and frequency comb generation in cavity-enhanced second-harmonic generation, Phys. Rev. Lett. 116 (2016) 033901.

[55] N. Chauhan, A. Isichenko, K. Liu, J. Wang, Q. Zhao, R.O. Behunin, P.T. Rakich, A.M. Jayich, C. Fertig, C. Hoyt, others, Visible light photonic integrated Brillouin laser, Nat. Commun. 12 (2021) 4685. https://doi.org/10.1038/s41467-021-24926-8.

[56] M. Nie, K. Jia, J. Bartos, S. Zhu, Z. Xie, S.-W. Huang, Turnkey photonic flywheel in a Chimera cavity, ArXiv Prepr. ArXiv221214120 (2022).

[57] M. Nie, S.-W. Huang, Quadratic soliton mode-locked degenerate optical parametric oscillator, Opt. Lett. 45 (2020) 2311–2314.

[58] D. Zhu, L. Shao, M. Yu, R. Cheng, B. Desiatov, C.J. Xin, Y. Hu, J. Holzgrafe, S. Ghosh, A. Shams-Ansari, E. Puma, N. Sinclair, C. Reimer, M. Zhang, M. Lončar, Integrated photonics on thin-film lithium niobate, ArXiv210211956 Phys. (2021). http://arxiv.org/abs/2102.11956 (accessed March 2, 2021).

[59] L. Yin, J. Zhang, P.M. Fauchet, G.P. Agrawal, Optical switching using nonlinear polarization rotation inside silicon waveguides, Opt. Lett. 34 (2009) 476–478. https://doi.org/10.1364/OL.34.000476.

[60] R. Pant, C.G. Poulton, D.-Y. Choi, H. Mcfarlane, S. Hile, E. Li, L. Thevenaz, B. Luther-Davies, S.J. Madden, B.J. Eggleton, On-chip stimulated Brillouin scattering, Opt. Express 19 (2011) 8285–8290. https://doi.org/10.1364/OE.19.008285.

[61] A. Andrushchak, B. Mytsyk, H. Laba, O. Yurkevych, I. Solskii, A. Kityk, B. Sahraoui, Complete sets of elastic constants and photoelastic coefficients of pure and MgO-doped lithium niobate crystals at room temperature, J. Appl. Phys. 106 (2009). https://doi.org/10.1063/1.3238507.